\documentstyle[psfig]{lamuphys}
\makeatletter
\let\chapter\hid@chapter
\makeatother
\renewcommand{\thefootnote}{\fnsymbol{footnote}}
\begin{document}
\pagenumbering{arabic}
\title{Evolution of Elliptical Galaxies up to $z\approx 1$ $^*$}

\author{R. Bender, R.P. Saglia, B. Ziegler}

\institute{Universit\"ats--Sternwarte, Scheinerstra\ss e 1,
D--81679 M\"unchen, Germany}

\maketitle

\begin{abstract}
We review the observational evidence showing that luminous cluster 
elliptical galaxies are old stellar systems, undergoing mostly passive stellar
evolution up to redshift $z\approx 1$, with approximate coeval epoch
of formation. This scenario is supported by observations of local
early--types, collected by 2m -- 4m class telescopes, and fits the recently
gained high resolution imaging (given by the refurbished HST) and deep 
spectroscopic data coming from 4m telescopes of $z\approx 0.4$ cluster
ellipticals. Up to $z\approx 1$, luminosity functions, colors and
surface brightnesses provide further evidence for mild evolution of
massive cluster ellipticals.

The new 8m class telescopes and in particular the VLT with its imaging and
multiplex spectroscopic instruments will give further essential
information on early--type galaxy evolution over the full galaxy
mass--spectrum and as a function of environment, disk--to--bulge ratio
and other parameters. The VLT will also allow to explore the
possibility to use evolution--calibrated massive ellipticals as
cosmological standard candles/rods.

\end{abstract}

\footnotetext[1]{\sc Review presented at the ESO Workshop ``The Early Universe
with the VLT'', eds. J. Bergeron et al., Springer Verlag, Berlin, 1996}

\renewcommand{\thefootnote}{\arabic{footnote}}

\section{Introduction}

What was the major epoch of galaxy formation? How extended is this
epoch?  Are elliptical galaxies old? When did the last episodes of
star formation happen in early--type galaxies? Can elliptical galaxies
be used as cosmological standard candles? These are some of the most
discussed questions in the last 20 years of studies of early--type
galaxies. A coherent picture is now slowly emerging, which may be able to
answer these questions and trace the history of star formation in
these galaxies.

In Section 2 we review formation scenarios and age estimates for elliptical
galaxies based on the observations of local field and cluster samples. In
Section 3 we discuss the constraints on the evolution of the stellar
populations of ellipticals up to redshifts $z\approx 1$ coming from the
study of their colors, luminosity function, surface brightness, Fundamental
Plane and ${\rm Mg}-\sigma$ relation and we illustrate how cluster
ellipticals can possibly be calibrated and used as standard candles. In
Section 4 we highlight a number of projects which involve the study of the
properties of early--type galaxies at high redshift and which will be best
performed with the VLT and its imaging and spectroscopic instruments.  In
Section 5 we summarize our conclusions.

\section{2m Science: constraints from local E/S0}

The large amount of observational material accumulated over the last
decade with medium--class telescopes (2m -- 4m telescopes) allows one
to put some important constraints on the merging and star formation
(SF) history of early--type galaxies in the "local" universe.

About 1/3 of luminous nearby ellipticals show kinematically decoupled
cores or otherwise peculiar kinematics. Such configurations point to
formation scenarios which involve major merging events between
progenitor objects with low gas--to--star ratio (see, e.g.,
Bender 1996, Barnes 1996). In addition, the statistics of
counter--rotating gaseous
disks present in E and S0 galaxies indicate that each early--type galaxy
experiences accretion of a low luminosity galaxy or of intergalactic gas
at least once in its life time (Bertola et al. 1992). Finally, the
presence of shells, tails, x--shaped structures etc., mostly in field
galaxies, is further indication for at least  minor merging or
accretion events taking place today in low density environments
(see review by Schweizer 1990).

On the other side, the small scatter observed in the colour--velocity
dispersion ($\sigma$) and ${\rm Mg}_b-\sigma$ relations indicates that the
bulk of the stars of {\it luminous cluster} ellipticals must be rather old
and must have an almost coeval epoch of formation, with $\Delta t/t
\la 0.15$ (Bower et al.  1992; Bender, Burstein, Faber 1993), in
agreement with recent calculations in the context of CDM models
(Kauffmann 1996). The high ${\rm Mg}_b$ absorption values and the large
$[{\rm Mg}/{\rm Fe}]$ ratios imply also large ages and presumably short star
formation time scales for luminous Es (Matteucci 1994, Bender
1996, Greggio 1996). In addition, the small scatter observed in the
mass--to--light ratio ($M/L$) derived from the "Fundamental Plane"
analysis sets tight constraints on the relative variations in
dynamics, IMF and ages of the stellar populations (Renzini \& Ciotti
1993). Finally, the absorption--index diagrams between H$_\beta$, Fe
and ${\rm Mg}_b$ confirm the old age for luminous Es, but imply smaller
ages or extended SF histories for low luminosity early--type galaxies
(Faber, Worthey, Gonzalez 1992, Gonzalez 1993). 
Since most low luminosity ellipticals
are likely to contain disks contributing up to 30\% to the total
light (Bender et al. 1989, Rix and White 1990, Scorza and Bender
1995), there is still the possibility that younger {\it mean}  ages may
solely be associated with extended star formation in disks while
bulges may still be old. Some low luminosity Es (especially compact
Es) may also be "disk--less" bulges or bulges with stirred disks (see
Bender et al. 1992). Independent from this uncertainty, ellipticals in
low density
environments show more
peculiarities and may be genuinely younger than cluster ellipticals
(Schweizer et al. 1990, Kauffmann 1996,  Gonzalez
1993), though the evidence is conflicting (De Calvalho \& Djorgovski 1992,
Lucey \& Guzm\'an 1993, Burstein 1989).

Summarizing, the above observational facts suggest the following
conclusion.  Luminous cluster ellipticals formed rapidly in early
merging events. Luminous field ellipticals may be younger and could be
late mergers. Faint ellipticals and S0 galaxies may have had extended
SF histories. 

\section{4m Science $+$ HST: redshift evolution of elliptical galaxies}

Confirmation and even tighter constraints on stellar population ages 
in early--type galaxies come from the combined use of 4m
(spectroscopic) telescopes and the imaging capabilities of the
refurbished HST. In order of increasing accuracy, the following tests
of the evolutionary history of early--type galaxies can be listed.

The median colours of cluster E galaxies evolve only slowly with
redshift, consistent with mostly passive evolution up to $z\approx 1$
(Aragon--Salamanca et al. 1993, Stanford et al. 1995).  The galaxy
counts, divided as a function of morphological type, show that the
number density of E/S0 galaxies does not evolve with redshift within
current errors (Driver et al. 1996). The luminosity function of red
galaxies and K--band selected galaxies changes only very weakly with
redshift up to $z\approx 1$ (Glazebrook et al. 1995, Lilly et
al. 1996, Ellis et al. 1996).

The surface brightnesses of E/S0s decrease with redshift following
closely the Tolman relation ($I\approx (1+z)^{-4}$) and passive
evolution models (Franx 1993, 1995; Dickinson 1995, Pahre et
al. 1996), up to redshifts $z\approx 1$.

More recently, pushing to the limit the current generation of 4m
telescopes and instrumentation, it has been possible to investigate
problems that will be best tackled with the future 8m telescopes. The
evolution of the Fundamental Plane (and the related $M/L$) has been
followed up to $z\approx 0.4$. It is consistent with mostly passive
evolution, but smaller SF events cannot be excluded (Franx 1993, 1995,
see also this conference).  

On the same line of research, the evolution of
the ${\rm Mg}-\sigma$ relation (Bender, Ziegler, Bruzual 1996)
demonstrates that luminous cluster ellipticals are very old. Figure 1
shows the ${\rm Mg}_b-\sigma$ relation for the clusters A370 and
MS1512+36, at redshift $z\approx 0.375$, together with the
local ellipticals of the Virgo and Coma cluster.  An aperture
correction has been applied to put the two samples on the same scale. 
Distant ellipticals also show a correlation between ${\rm Mg}_b$ and
$\sigma$ as local ellipticals do. However, there is clear evidence for
evolution: at any given $\sigma$, the mean ${\rm Mg}_b$ of Es at $z=0.375$ is
lower than at $z=0$. The evolution is very small for massive Es and
likely stronger for faint Es, on average it is about 0.3 \AA. Using Worthey's
(1994) population synthesis models, the ${\rm Mg}_b$ weakening at a given
$\sigma$ can be translated into a relative age difference. This
results in the conclusion that the bulk of the stars in the luminous
cluster ellipticals must have formed at redshifts $z>2$. Moreover, the
same models can be used to translate the observed ${\rm Mg}_b$ weakening
into a --0.4 mag change of the B--band luminosity, assuming a Salpeter IMF. 
The fact that the slope of the ${\rm Mg}_b-\sigma$ relation at
$z=0.375$ appears to be slightly steeper than today
indicates that less luminous ellipticals may be generally
younger (consistent with Faber et al. 1992 local H$_\beta-{\rm Mg}-$Fe
measurements, see previous section). Note that, in contrast to luminosity
and surface brightness evolution tests, the ${\rm Mg}_b-\sigma$ test
is independent of the slope of the IMF in elliptical galaxies (see
discussion in Bender et al. 1996).

We have imaged some of the ellipticals observed spectroscopically by Bender et
al. (1996) with HST and determined the structural parameters $R_e$
(the half--luminosity radius) and $\langle SB_e\rangle$ (the effective
average surface brightness) in the F675W filter, using the
two--component fitting algorithm developed by Saglia et al. (1996),
with HST psf convolution tables.  The surface brightness term has been
transformed to rest--frame B--band,
corrected for cosmological dimming ($(1+z)^4$) and for passive
evolution using the ${\rm Mg}_b-\sigma$ relation described above. 
Figure 2 shows the Fundamental Plane of the clusters A370 and
MS~1512+36 (filled/open squares before/after evolution correction)
together with the data points of the Coma ellipticals (open circles),
all having $\sigma>150$ km/s. The distances are computed using
$H_0=50$ km/s/Mpc and $q_0=0.25$. The figure shows that the passive evolution
correction fully explains the observed luminosity evolution and allows
for $q_0$ values in the range $0-0.5$.

\begin{figure}
\centerline{\psfig{figure=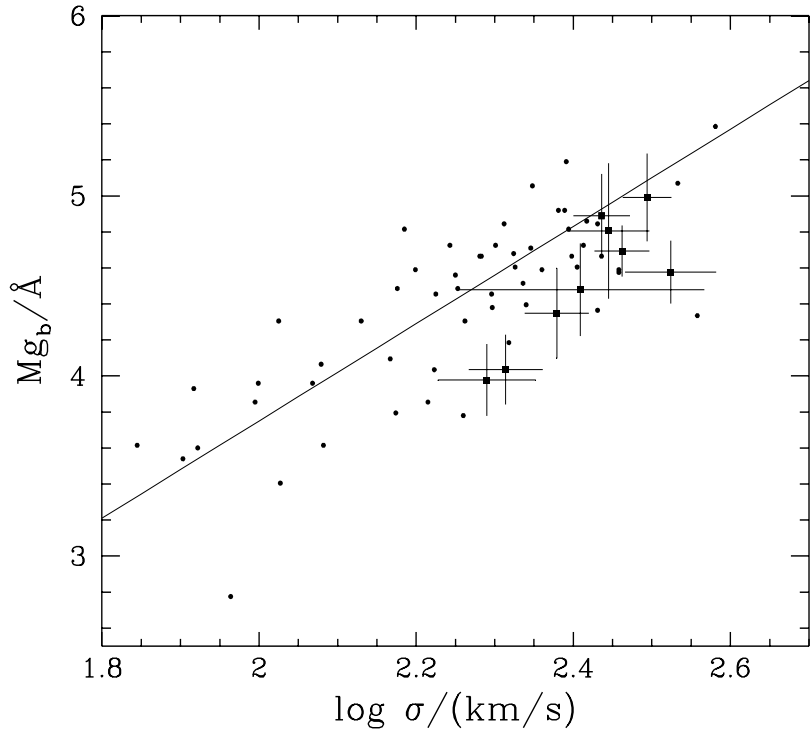,height=11cm,width=14cm,angle=-90}}
\caption[*]{The ${\rm Mg}_b-\sigma$ relation  for ellipticals in  the
clusters A370 and MS1512+36 at redshift $z=0.375$ (filled squares), 
compared to the one of Coma Cluster ellipticals (small filled  circles).}
\end{figure}

\begin{figure}
\centerline{\psfig{figure=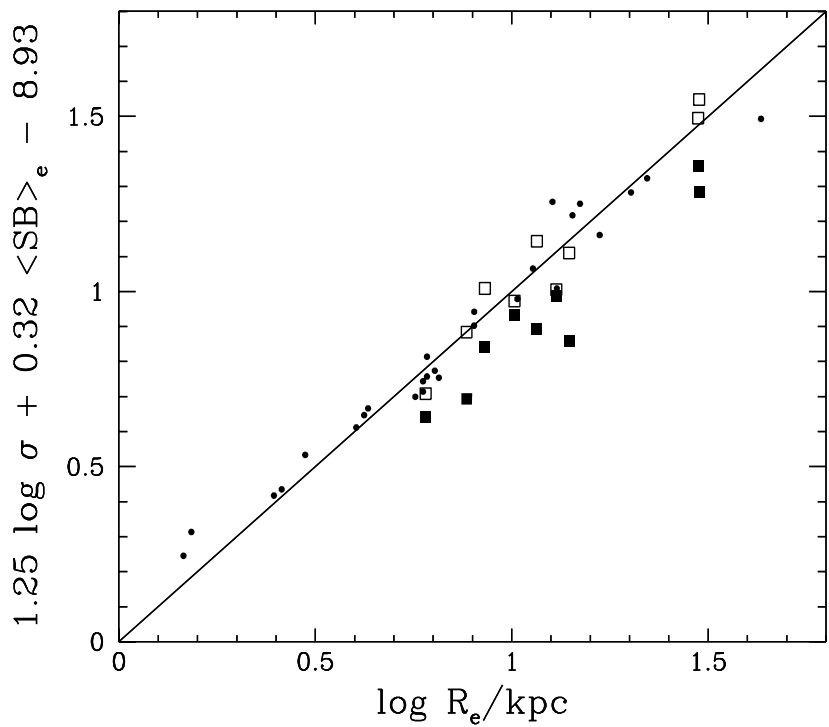,height=11cm,width=14cm,angle=-90}}
\caption[*]{The Fundamental Plane at redshift $z=0.375$ for the same 
elliptical galaxies as in Figure 1 (filled/open
squares: before/after correction for luminosity evolution). 
The filled small circles show the position of Coma Cluster
ellipticals. We used $H_0=50$ km/s/Mpc and $q_0=0.25$.}
\end{figure}

\bigskip
In contrast to cluster ellipticals, blue members and E$+$A galaxies
are known to show strong evolution (Butcher and Oemler 1978,  Dressler and
Gunn 1983). However, these objects are unlikely to end as ellipticals.
Hubble Space Telescope imaging of blue and E$+$A galaxies
indicates that presumably most of them are infalling spirals
experiencing tidal shaking or 'harrassment' (Moore, Katz, Lake 1996)
with only a small percentage undergoing merging events. Since the
outer parts of the disks are stripped during these processes and/or
star formation is likely to enhance the central stellar densities,
large disk--to--bulge ratios are transformed into low disk--to--bulge
ratios. The end products of this process are therefore likely to
resemble S0 galaxies or disky ellipticals. In fact, S0s, and not
ellipticals, are the dominant galaxy population in clusters at lower
redshifts (e.g., Saglia et al. 1993, J\o rgensen et al. 1994).

Summarizing, the conclusions based on the redshift evolution of
elliptical galaxies are in full agreement with the analysis of the
properties of local samples of ellipticals (see Section 2).

\section{8m Science (+HST)}

Following the discussion of the previous sections, it is natural to
highlight a few tasks that the VLT (and other 8m telescopes) will
be able to perform (the list is of course very incomplete). 

The search for the oldest ellipticals at high redshifts is connected
to the determination of the epoch of first star formation, a crucial
test of cosmological models. Some of these objects may have been found
already (see Giavalisco, this conference), but larger statistics and
better information about these objects is
still needed. As preparatory work to the
spectroscopic observations with the VLT, a (photometric) survey of
distant clusters ($z\ga 0.6$) in the southern hemisphere of the kind
described by Da Costa et al. (1996) is needed.

An extremely powerful test of cosmological models of structure
formation is the determination of the evolution of the {\it potential
function} (or {\it dispersion function}) of E, S0 and spiral galaxies
via the measurement of their internal kinematics as a function of
redshift. This will allow to follow the evolution of the potential
depth of dark matter halos directly, bypassing the uncertain steps
related to the ill--known processes of star formation needed to compute
the luminosity function.

The VLT, in combination with HST photometry, can allow the accurate
determination of the redshift evolution of the luminosities, stellar
population parameters, internal kinematics and structural parameters
of galaxies.  This will be the essential information to test galaxy
formation models in detail, i.e., beyond dark halo evolution. 
In addition, this information can possibly be
used to calibrate elliptical galaxies as standard candles and
cosmological tools, to be used to constrain $q_0$ via the Fundamental
Plane relations and the Volume test.

Spectroscopic instruments with multiplex capabilities and low/medium
resolution in the optical and near infrared spectral range such as
FORS, ISAAC, NIRMOS/VIRMOS will play a decisive role in the above
outlined projects. The imaging capabilities offered by FORS and
especially CONICA plus Adaptive Optics in the NIR will complement the
HST high resolution with the power of collected flux.

\section{Conclusions}

Observations of local massive ellipticals obtained with 2m -- 4m class
telescopes over the last decade demonstrated that these objects, though most
likely being formed in mergers, exhibit only small scatter in 
color/line--strengths vs. velocity--dispersion diagrams and in
mass--to--light ratios (from 'fundamental plane' analysis).
The small amplitude of the scatter indicates {\it approximately 
coeval formation of massive ellipticals at high redshift}.  Further
support for this conclusion comes from H$_\beta$ vs.
metal--linestrength analysis and from the high overabundance of light
elements relative to iron (indicating short star formation time
scales). Age and formation constraints on lower luminosity early--type
galaxies and also field ellipticals are less tight and lower mean ages
or extended star formation histories are possible.

Observations of redshifted luminous cluster ellipticals with 4m--class
telescopes and the Hubble Space Telescope up to $z \approx 1$ confirm
this picture. Luminosities, surface brightnesses and mass--to--light
ratios, as well as the fundamental plane relation evolve only slowly
with redshift $z$. The most reliable measurements show a rest--frame
B--band evolution following the simple relation $\Delta B(z) \approx -z$.
Again, our information on low luminosity ellipticals and field
ellipticals at higher redshift is much more uncertain.

These results put strong constraints on the formation of luminous
cluster ellipticals and may further offer a route to calibrate these
objects as cosmological standard candles or standard rods.

\bigskip
{\bf Acknowledgments:} Financial support by the Deutsche
Forschungsgemeinschaft under SFB 375 is gratefully acknowledged.
We thank our collaborators Drs. Paola Belloni, Laura Greggio and Ulrich
Hopp for discussions and comments. Part of the work presented here is
based on observations made with the NASA/ESA Hubble Space Telescope,
obtained at the Space Telescope Science Institute, which is operated
by the Association of Universities for Research in Astronomy, Inc.,
under NASA contract NAS 5-26555.

%
%
%

\end{document}